
\MAINTITLE={The proton blazar
}
\AUTHOR={K.~Mannheim}
\INSTITUTE={Max--Planck--Institut f\"ur Radioastronomie\newline
Auf dem H\"ugel 69, W--5300 Bonn 1\newline
Federal Republic of Germany
}
\RECDATE={July 8, 1992}
\ACCDATE={November 8, 1992}
\SUMMARY=
{
Recent gamma ray detections and the radio/X-ray correlation of extragalactic
flat-spectrum radio
sources make the existence of a ultra--relativistic proton
population in jets very probable.  The
protons with maximum
Lorentzfactors in the range $10^9-10^{11}$
generate hard photons with energies from
keV to TeV via
pion and pair photoproduction and subsequent synchrotron
cascade reprocessing.

In this paper relativistic
protons are considered in the context of
the Blandford and K\"onigl (1979) model of compact
radio jets which are assumed to be
responsible for the nonthermal
emission component of flat--spectrum quasars and
BL~Lacertids.  The baryons are simply added  to the
electrons considered as the only
radiative agent in the original work.

The
differential gamma ray spectrum induced by the protons
is an inverse power--law with an index
preferentially in the range $1.8$ to $2.0$.  Below a few MeV
the spectrum flattens to an X--ray spectrum with index $1.5$
to $1.7$.
For a given
energy flux through a jet the
apparent radio luminosity is lower for a
relativistic proton
enriched plasma than for a relativistic
electron enriched plasma ({\it i.e.} for jets
with a proton/electron energy density ratio
$\eta=u_{\rm p}/u_{\rm e} > 1$), while the apparent
gamma ray luminosity is almost constant.

Differential
Doppler--boosting, {\it i.e.} changes in flux
amplification due to changes in the orientation
of the radiating plasma element, influences the
pair creation optical depth and therefore has an effect
on the maximum photon
energies of the compact radio source. Hence TeV emission should
correlate with high states of the entire continuum
flux.

Further
consequences are
a)  an observable diffuse flux of neutrinos with a flat
spectrum in the PeV--EeV range,
b)  a diffuse flux of
gamma rays equal to the flux of gamma
rays from the Milky Way at an energy of 7~TeV,
c)  neutrons escaping the blazar that convert their
luminosity in the kinetic power of a conical wind surrounding the
jet or escape the host galaxy at highest energies.
}
\KEYWORDS={Elementary particles -- BL~Lacertae objects: general --
Galaxies: jets -- Gamma rays: theory -- X--rays: galaxies}
\THESAURUS={02.05.1, 11.02.1, 11.10.1, 13.07.3, 13.25.2}
\SENDOFF={K. Mannheim}
\maketitle
\titlea{Introduction}
Extragalactic radio sources are the largest known dissipative
structures (nonthermal objects) known in the universe and as such highly
interesting candidate sources of
cosmic rays (see references in Berezinski\u i et al. 1990).
There are two ways for them to inject cosmic ray baryons
into the intergalactic medium; (i) by escape from hot spots often a
few hundred kiloparsecs away from active nucleus of the host galaxy
(Biermann 1991,
Rachen and Biermann 1992) and (ii) by neutron
escape from the sub--parsec scale of the jet
or the nucleus
(Protheroe and Szabo 1992).  Both these mechanisms avoid adiabatic losses
preventing such escape otherwise.   Do we have any indications
that baryons are actually present in radio jets?

The leptonic component ($e^\pm$) of nonthermal particles in
jets are the well--studied origin of synchrotron radiation
from radio to X-ray frequencies ({\it e.g.},  see the
conference proceedings edited by Maraschi et al. 1989 and
Zensus et al. 1987, Bregman 1990).  Subtracting thermal emission
components like dust infrared radiation and the big blue bump in
quasars one is left with a continuum spectrum of highly varying
flux and polarization.  Flat--spectrum radio sources are
characterized by jets oriented at small angles to the line of sight, so
that the radiation from the base of
the jet  is Doppler--boosted towards the observer.
Some of these sources, the blazars, show
very active behaviour of the most
compact regions of the jet, especially
with respect to polarization.
Probably all flat--spectrum radio sources appear as blazars during
active episodes.
To simplify things the
compact jet with relativistic electrons and protons shall be
coined the ``proton blazar'' throughout this work.

Since the fact that particle
acceleration takes place in radio jets is obvious from
synchrotron observations, it is a small speculative step to argue
that this acceleration mechanism not only concerns electrons
(and positrons), but protons (and nuclei)
as well.   However, the step is really not speculative, because
it has already  been shown that in principle
the gamma radiation from flat--spectrum
radio sources detected by GRO can be explained as
proton initiated synchrotron cascade emission  in radio jets
(Mannheim and Biermann 1992) assuming that protons are shock
accelerated up to energies of $10^{11}$~GeV.
The cascade emission ranges
from X-rays to gamma rays  competing with
synchrotron--self--Compton (SSC) emission (K\"onigl 1981).
The cascade spectrum
is always harder than $\alpha_X=1$, so that the luminosity of this
mechanism peaks in the gamma ray band, unless
there is thermal reprocessing  (Zdziarski et al. 1990).
This raises the question of how
far one can actually interpret  nonthermal emission above keV as
proton induced and what properties determine the high energy
blazar spectrum.   The physical conditions inferred may -- among
other things -- help
to elucidate the role of extragalactic jets as sources of cosmic rays.

To this end one must specify the proton and electron distributions in
a physically constrained plasma volume element.  The Blandford
and K\"onigl (1979) model of compact relativistic jets shall serve
as a conceptual guideline here (BK model hereafter).
Models of shock acceleration explain how particle acceleration in
jets possibly works ({\it e.g.}, Ellison et al. 1990).
One important cooling channel of the protons is
photomeson (mainly pion) production  (Sikora et al. 1987).   The pions
decay yielding
neutrinos, pairs and gamma rays.
Both, the weak and the electromagnetic fraction of the pion
power, have very important observational consequences.

Neutrino detection is feasible
with  experimental
designs currently under construction like AMANDA, DUMAND II
or BAIKAL (Stenger et al. 1992) in the TeV range,
by advanced analysis of horizontal atmospheric showers up
to the PeV range (Halzen and Zas 1992) and
at still higher energies with the High Resolution Fly's Eye (Cassiday et
al. 1989).
Extragalactic neutrino beams would be a unique tool for
elementary particle physics to
study weak interactions in a kinematical regime far beyond planned laboratory
designs.

In contrast to the neutrinos the
electromagnetic power  injected into the compact
jet at energies far
above TeV is observable only through the action of several reprocessing
cycles with pair creation followed by synchrotron radiation.
The target photons are again, self--consistently as for the protons, the soft
synchrotron photons from the primary electrons.
Such electromagnetic showers  redistribute the power down below the
critical energy $E_\gamma^*$ where the compact jet is
optically thick to pair creation (Burns and Lovelace 1982,
Svensson 1987, Berezinski\u i et al. 1990 and references therein).
In the BK model it turns out that the typical energy (as seen in
the observer's frame) is in the TeV range.  The showers are
therefore unsaturated in the sense that not all photons with
sufficient energy to create a pair (with energy above
twice the electron rest mass) do actually create a pair.  As a consequence
a detectable annihilation line does not form.

The critical energy is commonly expressed in terms of the radiation compactness
$l$,
which is proportional to the energy--dependent
pair creation optical depth $\tau_{\gamma\gamma}(E_\gamma)\simeq
0.1 l E_\gamma /m_{\rm e}c^2$
at  electron rest mass energy.
Assuming isotropic emission in the comoving frame of a  relativistic jet the
target radiation compactness  is given by
$$\eqalign{l&\simeq
{\sigma_{\rm T}\over 4\pi m_ec^3}{L_{\rm s}\over r'}\simeq
2.1\times 10^{-30}{L_{\rm s}\sin\theta\over r_{\rm ob}\gamma_{\rm j}}\simeq
10^{-5}\cr &\times
\left[{L_{\rm s}\over 10^{43}\thinspace
{\rm erg/s}}\right]
\left[\gamma_{\rm j}\over 10\right]^{-1}
\left[{r_{\rm ob}\over 10^{-2}\thinspace{\rm pc}}\right]^{-1}
{\sin\left[\theta\right]\over\sin\left[10^o\right]}\cr}
\eqno(1)$$
where the comoving frame synchrotron luminosity is given by
$$L_{\rm s}=D_{\rm j}^{-p}L_{\rm (obs)}\eqno(2)$$
with $p=4$ (Urry and Shafer 1984), the observed
source luminosity is $L_{\rm (obs)}$ and the Lorentzfactor, speed  of the
jet and angle to the line of sight are $\gamma_{\rm j}$, $\beta_{\rm j}$
and $\theta$, respectively.
Here
$$D_{\rm j}=\left[\gamma_{\rm j}(1-\beta_{\rm
j}\cos\theta)\right]^{-1}\eqno(3)$$
denotes
the Doppler--factor of the jet which can be estimated from
superluminal expansion of VLBI knots yielding
near maximum flux amplification $D_{\rm j}\approx
\gamma_{\rm j}\approx 2-10$, so that
$L_{\rm s}\ll L_{\rm (obs)}$.
The length $r$ of the
conical jet projected on the line of sight is given by
$r_{\rm ob}=r\sin\theta$ and the intersection of the l.o.s.
with the jet in the comoving frame is given by $r'=
r_{\rm ob}\gamma_{\rm j}/\sin\theta$
provided the bright part of the  jet is viewed along its entire length.

{}From Eq.(1) one obtains
for a target spectral index $\alpha=1$ the critical energy in the
observer's frame
$$\eqalign{&E_\gamma^*\simeq 10D_{\rm j}
l^{-1}\thinspace{\rm MeV}
 \simeq 10{\rm \thinspace TeV}\cr & \times
\left[{L_{\rm s}\over 10^{43}\thinspace
{\rm erg/s}}\right]^{-1}
\left[{r_{\rm ob}\over 10^{-2}\thinspace{\rm pc}}\right]
\left[{\gamma_{\rm j}D_{\rm j}\over 100}\right]
\left[\sin\left[\theta\right]\over \sin\left[10^{\rm o}\right]\right]^{-1}
\cr}\eqno(4)$$
Thus gamma rays even up to 10~TeV can be emitted by
blazars in detailed accordance with recent GRO (Hartman et al. 1992) and
Whipple
(Punch et al. 1992)
observations.
The fact that
$L_{\rm s}\ll L_{\rm (obs)}$
is actually a nice property in the framework of unifying models,
where flat--spectrum radio quasars are assumed to
be
Fanaroff--Riley class II (FR~II) galaxies (Fanaroff and Riley 1974)
with the jet viewed at small angles to the line of sight.
In this case
the radiative dissipation in the nuclear jet must  be only marginal --
otherwise the core would not be able to sustain the large--scale
jet of the radio galaxy with its enormous kinetic luminosity.
The kinetic luminosity of the jet must exceed
the synchrotron luminosity  $L_{\rm s}\approx 10^{44}$~erg/s
of hot spots at the ends of FR~II galaxies.

The organization of the paper is as follows:
Firstly,  properties of a typical synchrotron component
in the BK model
are described including relativistic
electrons and protons.  Results are then compared to
observed properties of 3C279 and Mkn421.
Hereby I
demonstrate the action of parameters involved with the model.
Secondly, an estimate of the gamma ray and neutrino
background contribution from proton blazars
is obtained.  Finally, the consequences of biconical escape
of neutrons are studied.

\titlea{A rudimentary proton blazar model and its consequences}
In this section I refer to
the explanation of flat--spectrum radio sources as compact relativistic
jets  demonstrated in the model of Blandford and K\"onigl (1979)
by simply {\it adding}
a population  of accelerated protons.
The protons  cool in the synchrotron photon
atmosphere generated by the accelerated electrons.

The striking discovery of intraday radio variability
emphasizes the fact that shocks in compact jets may
indeed be the accelerating agent of blazars
(Witzel 1990).  As a matter of fact, the extremely short
time scale of the variations can be explained best by relativistic shocks
propagating down a relativistic bulk flow with embedded inhomogenities
(Qian et al. 1990).
When the trajectories of the shocks
are not linear, but helical
as suggested by VLBI pictures
or when the distribution of inhomogenities is
ordered, the resulting variability pattern may even be quasi--periodic as
observed in 0716+714 (Quirrenbach et al. 1991).
The  BK compact jet may therefore be considered a crude
approximation to
the downstream regions of multiple, probably oblique and relativistic
shocks propagating through a jet flow coming out of an AGN.

\titleb{ The model}
In the Blandford and K\"onigl model it is assumed that electrons are
accelerated in a supersonic conical jet of
opening angle $\Phi$ yielding
a synchrotron spectrum
locally with the optical thin index $\alpha=0.5$
up to the break frequency
determined by equal loss and expansion time scales, respectively, {\it viz.}
$$t_{\rm e}=7.7\times 10^8\gamma_{\rm j}
B^{-2}\gamma_{\rm e}^{-1}\ {\rm s}\eqno(5)$$
and
$$ t_{\rm exp}=10^8r_{\rm ob}(\sin\theta)^{-1}\beta_{\rm j}^{-1}\ {\rm
s}\eqno(6)$$
where $B=B_1(r/1\thinspace{\rm pc})^{-1}$ denotes the magnetic
field strength in the jet (in Gauss).
Above the break frequency (in the observer's frame and in units of GHz)
$\nu_{b9}=4.2\times 10^{-3}(1+z)^{-1}D_{\rm j}B\gamma_{\rm e,b}^2$
the spectrum is loss dominated, so that
$\alpha=1$ up to the maximum frequency determined by equal acceleration
and loss time scale (Biermann and Strittmatter 1987) provided the acceleration
time scale remains shorter than the expansion time at the maximum energy.
Corresponding particle distributions break from $dN/dE\propto E^{-2}$
to $dN/dE\propto E^{-3}$.
The dimensionless break energy of the electron distribution is given by
$$\gamma_{\rm e,b}=7.7\gamma_{\rm j}\beta_{\rm j}
B_1^{-2}r_{\rm ob}(\sin\theta)^{-1}
\eqno(7)$$
Inserting the magnetic field strength
$$B_1=0.04
\Delta^{-{1\over 2}}
\left(1+{2\over 3}k_{\rm e}\Lambda_{\rm e}\right)^{-{1\over 2}}
\gamma_{\rm j}^{-1}\beta_{\rm j}^{-{1\over 2}}
(\sin\theta)^{-1}
\Phi_{\rm ob}^{-1}
L_{44}^{1\over 2}
\eqno(8)$$
at the radius $r_{\rm b,ob}$ (in parsec) where
the maximum brightness temperature is achieved
$$\eqalign{&r_{\rm b,ob}=0.04k_{\rm e}^{1\over 6}
\Delta^{-{13\over 12}}\left(1+{2\over 3}k_{\rm e}\Lambda_{\rm
e}\right)^{-{13\over 12}}
\gamma_{\rm j}^{-{19\over 6}}\beta_{\rm j}^{-{25\over 12}}
D_{\rm j}^{-{1\over 6}}\cr &\times
(\sin\theta)^{-{7\over 6}}
\Phi_{\rm ob}^{-2}
L_{44}^{13\over 12}\cr}\eqno(9)$$
yields the electron break energy
$$\eqalign{&\gamma_{\rm e,b}=200
k_{\rm e}^{1\over 6}
\Delta^{-{1\over 12}}
\left(1+{2\over 3}k_{\rm e}\Lambda_{\rm e}\right)^{-{1\over 12}}
\gamma_{\rm j}^{-{1\over 6}}
\beta_{\rm j}^{-{1\over 12}}
D_{\rm j}^{-{1\over 6}}\cr &\times
(\sin\theta)^{-{1\over 6}}L_{44}^{1\over 12}\cr}
\eqno(10)$$
Here the notation is the same as in Blandford and K\"onigl (1979), {\it i.e.}
$\Phi_{\rm ob}=\Phi/\sin\theta$,
$L_{44}$ denotes the luminosity of the jet
due to the energy flux in relativistic particles and fields in units of
$10^{44}$~erg/s, $\Delta=\ln r_{\rm max}/ r_{\rm min}\approx 5$,
$\Lambda_{\rm e}=\ln\gamma_{\rm e,max}/\gamma_{\rm e,min}\approx 4$.
One important consequence of including  protons is, however,
that $k_{\rm e}^{\rm BK}$ defined by
$k_{\rm e}^{\rm BK}\Lambda_{\rm e}=u_{\rm e}/u_B$ where
u denotes energy density
has to be replaced by
$$k_{\rm e}=k_{\rm e}^{\rm BK}
\left(1+\eta
\right)^{-1}\eqno(11)$$
where the energy density ratio of protons and electrons
is given by
$$\eta={u_{\rm p}\Lambda_{\rm p}\over u_{\rm e}
\Lambda_{\rm e}}\eqno(12)$$
In order to obtain the break energy of the proton distribution one must
equal the expansion time scale with the energy loss time scale for protons.
The ratio of electron and proton cooling time scales is given by
$${t_{\rm p}(\gamma_{\rm e})\over t_{\rm e}(\gamma_{\rm p})}
=\left[m_{\rm p}\over m_{\rm e}\right]^3\left[\gamma_{\rm e}\over
\gamma_{\rm p}\right]\left[1+a_{\rm s}\over 1+240a_{\rm s}\right]\eqno(13)$$
where the ratio of energy in synchrotron photons to the magnetic field
is denoted
$$a_{\rm s}={u_{\rm s}\over u_B}\approx k_{\rm e}\beta_{\rm j}\gamma_{\rm j}
\Phi\eqno(14)$$
The electron cooling time scale involves synchrotron and first--order
self--Compton emission, the proton cooling time scale involves synchrotron
and pion production losses.  The Bethe--Heitler  pair luminosity is less than
50\% of the pion
luminosity due to the low value $\langle\kappa\sigma\rangle_{\rm e^\pm}
\approx 1$~$\mu$b peaking at proton rest frame photon energies of roughly
$10m_{\rm e}c^2$ compared to
$\langle\kappa\sigma\rangle_{\pi}\approx 50$~$\mu$b
at a threshold energy in the proton rest frame of $280m_{\rm e}c^2$.
Additional suppression results from the fact that target photons
close to the threshold energy are found
 at lower frequencies for
pair production than for pion production.  However, at lower
frequencies the target spectrum is flat ($\alpha=0$).
Compared to proton synchrotron radiation with characteristic
synchrotron photon
energy $E_{\gamma}({\rm p,syn})=1.6\times 10^{-17}
B_\perp \gamma_{\rm p}^2 \thinspace {\rm MeV}$
pion production wins
for values $a_{\rm s}>0.005$.
Hence putting $t_{\rm exp}=t_{\rm p}$ one obtains
$$\gamma_{\rm p,b}=\left[m_{\rm p}\over m_{\rm e}\right]^3
{1+a_{\rm s}\over 1+240a_{\rm s}}\gamma_{\rm e,b}
=6\times 10^9{1+a_{\rm s}\over 1+240a_{\rm s}}
\gamma_{\rm e,b}\eqno(15)$$
and then
$${L{\rm p}\over L_{\rm e}}\approx {u_{\rm p}\over u_{\rm e}}
{t_{\rm e}(\gamma_{\rm e,b})\over t_{\rm p}(\gamma_{\rm p,b})}
=\eta
\eqno(16)$$
Realizing that the proton induced luminosity dominates the hard X--
and gamma ray range, whereas the electron induced luminosity
dominates the radio to UV range, one obtains
$$L(>X)/L(<UV)\approx 1\ \ \ \
{\rm for}\ \ \ \  \eta=1.$$

However, when the acceleration time scale equals the expansion time
scale already at an energy $\gamma_{\rm p}^*<\gamma_{\rm p,b}$, the
proton distribution will be essentially cut off at $\gamma_{\rm p}^*$.
{\it E.g.}, the acceleration time scale for shock acceleration at
nonrelativistic shocks in fully developed Kolmogorov turbulence
given in Biermann and Strittmatter (1987) leads to this situation
in compact jets.  In this case the cooling time scales
of protons and electrons are not the same at their maximum (resp. break)
energies and hence the induced photon luminosities will only
be comparable, if a greater proton/electron ratio compensates for
the longer cooling time scale of the protons (Mannheim and Biermann 1992).
Another problem related to the extremely high
proton break energy Eq.(15) is that the model is close to
the limits imposed by the geometry of the source (Bell 1978).
The ratio of the
radius of gyration of a proton
$r_{\rm g}=10^{-12}B^{-1}\gamma_{\rm p}$
 at the break energy to the transverse
size of the jet $r_\perp\approx r_{\rm b, ob}\Phi_{\rm ob}$ at
the radius $r_{\rm b}$ of the maximum brightness temperature
is given by
$$\eqalign{&{r_{\rm g}(\gamma_{\rm p,b})\over r_\perp
(r_{\rm b,ob})}=
30
k_{\rm e}^{1\over 6}
\Delta^{5\over 12}
\left(1+{2\over 3}k_{\rm e}\Lambda_{\rm e}\right)^{5\over 12}
\gamma_{\rm j}^{5\over 6}
\beta_{\rm j}^{5\over 12}
D_{\rm j}^{-{1\over 6}}
\cr & \times
(\sin\theta)^{-{1\over 6}}
L_{44}^{-{5\over 12}}
{1+a_{\rm s}\over 1+240a_{\rm s}}
\cr}\eqno(17)$$
Note that extremely high proton energies $E_{\rm p}>10^9$~GeV
imply that the escaping flux of neutrons resulting from pion
production on the photon target $p+\gamma\rightarrow
n+\pi^+$ can escape the radio galaxy
and contribute after $\beta$-decay to the flux of cosmic rays observed at earth
(for cosmic rays of a neutron origin from
radio--quiet AGN cf. Protheroe and Szabo 1992).

A further important parameter is the compactness of the relativistic
jet.  It is calculated from Eq.(1) yielding
at the radius $r_{\rm b,ob}$
$$\eqalign{&l=9\times 10^{-4}
k_{\rm e}^{5\over 6}
\Delta^{13\over 12}
\left(1+{2\over 3}k_{\rm e}\Lambda_{\rm e}\right)^{1\over 12}
\gamma_{\rm j}^{13\over 6}
\beta_{\rm j}^{25\over 12}
D_{\rm j}^{1\over 6}
\cr &\times
(\sin\theta)^{13\over 6}
\Phi_{\rm ob}^2
L_{44}^{-{1\over 12}}
\cr}\eqno(18)$$
Eqs.(4), (15) and (18)  imply that the photons from pion decay
with mean energy at the maximum of the emissivity
$\langle E_\gamma\rangle\simeq 0.1 E_{\rm p,b}$
are fully reprocessed
by a synchrotron casacade $\gamma+\gamma_{\rm s}
\rightarrow e^++e^-$,
$e+\gamma^*\rightarrow e+\gamma$, $\gamma+\gamma_{\rm s}
\rightarrow e^++e^-$, etc. ($\gamma_{\rm s}$ denotes a soft
target photon and $\gamma^*$ denotes a virtual transversal photon
of the magnetic field),
until the electromagnetic power finally shows up below the TeV
range.  The cascade is started only because the pion flux is flat,
so that the injection of power has its maximum at the maximum
pion energy\footnote{$^{1)}$}{
A detailed description of
the physics of the proton initiated cascade
(PIC) is given in Mannheim et al. (1991),
hereafter refered to as MKB.  The reader may demonstrate him/herself
the action of the reprocessing cycles
by calculating back-of-the-envelope sequentially
$\pi^o$ decay photon injection,
photon-photon pair production, synchrotron radiation, pair production etc.
until the characteristic synchrotron frequency is below the value Eq.(4).}.
The dominant target photons for pion production are the photons
with $\nu>\nu_{\rm b}$, since they only satisfy the threshold condition
for pion production justifying the assumption $\alpha=1$ for the target
field (the target photons at threshold for pair production
have $\nu\ll \nu_{\rm b}$ reducing their contribution).

\titlec{The relative contribution of secondaries from pp and p$\gamma$
collisions, respectively}
Since the energy loss time scale for pp collisions is constant, in contrast
to the p$\gamma$ losses described above, the emissivity of secondaries
from pp collisions is much steeper, {\it i.e.} reflects the power
law of the protons.  This is especially  important
for the neutrino flux from proton blazars, which is flat for p$\gamma$
collisions, but has a $E^{-1}$ power--law for pp collisions.
Therefore the neutrinos from pp collisions are important at lower
neutrinos energies than the neutrinos from p$\gamma$ collisions.
In order to estimate the relative contributions one can readily calculate
the ratio $L_\nu({\rm pp})/L_\nu({\rm p}\gamma)$ of the neutrino
luminosity
(or photon luminosity analogously) induced by pp and p$\gamma$
collisions, respectively.

The thermal gas density in jets is limited by the fact that the kinetic
energy of the jet flow should remain less than or equal to the
Eddington luminosity of the AGN producing the jet.
$$L_{\rm kin}=An_{\rm th}m_{\rm p}c^3
\gamma_{\rm j}\beta_{\rm j}\le L_{\rm edd}=1.3\times 10^{46}
M_8{\thinspace\rm erg\thinspace s^{-1}}\eqno(19)$$
where $A=3.4\times 10^{34}r^2$~cm$^2$ denotes the cross sectional
area of the jet at length $r$ (in parsec) for an intrinsic
opening angle of $\Phi=2^{\rm o}$, $\gamma_{\rm j}\simeq 10$
and $M_8$ denotes the mass of
a central black hole in units of $10^8$ solar masses $M_\odot$.
{}From Eq.(19) one obtains
$$n_{\rm th}\le 7.9\times 10^{2}M_8
r^{-2}\thinspace {\rm cm^{-3}}\eqno(20)$$
Now the pp energy loss time scale is given by
$$t_{\rm pp}=\left[n_{\rm th}c\sigma_{\rm pp}\right]^{-1}\eqno(21)$$
with $\sigma_{\rm pp}\simeq 3\times 10^{-26}$~cm$^2$.
{}From Eqs.(5), (7) and (21) it follows that
$${L_\nu({\rm pp})\over L_\nu({\rm p\gamma})}={t_{\rm p\gamma}
(\gamma_{\rm p,b})
\over t_{\rm pp}}\le 7\left[r_{\rm S}\over r\right]\eqno(22)$$
where $r_{\rm S}\simeq 10^{-5}M_8$ denotes the Schwarzschild radius.
Thus even at the base of the jet, where $r\approx 10 r_{\rm S}$,
photon cooling is the dominant source of secondaries.

\begfig6cm
\figure{1}{The variation of the proton blazar
spectrum with the proton/electron ratio
$\eta$.
Parameters are $\theta=7^o$, $L_{44}=1$, $z=0.1$,
$\Delta=5$, $\Lambda_{\rm e}=4$, $\Lambda_{\rm p}=20$,
$k_{\rm e}^{\rm BK}=0.5$, $\gamma_{\rm j}=5$.
}
\endfig
\begfig6cm
\figure{2}{The variation of the proton blazar
spectrum with the angle to the line of sight $\theta$.
Parameters are as above with $\eta=1$.
}
\endfig

\titleb{The spectrum}
With the proton and electron distributions, as well as the magnetic
field and the compactness specified by the model one can calculate
the resulting PIC spectrum choosing a proper value of
the proton/electron ratio.  As a matter of fact, when gamma ray
observations have been performed one can estimate the proton/electron
ratio from $\eta=L(>X)/L(<UV)$, cf. Eq.(16).  The target spectrum
for the cascade is the integrated spectrum over the compact jet
which has the global spectral index $\alpha=0$ up to $\nu_{\rm b,ob}$
and then $\alpha=1$ up to the maximum primary electron
synchrotron frequency at $\nu_{\rm c}=10^{14-16}$~Hz.
Although the flatness of the radio spectrum follows from
the Blandford and K\"onigl model, the shock origin of the
accelerated particle populations emphasizes the importance of
a non--continuous distribution of surface brightness (separation
of shocks) and hence to a smoother transition $\alpha=0
\rightarrow \alpha=1/2\rightarrow \alpha=1$ to the loss
dominated spectrum.  {\it I.e.}, strong individual shocks
seemingly important in blazars where they cause the
extremely rapid variability may show up with their
optically thin index $\alpha=1/2$ at frequencies
just below $\nu_{\rm b,ob}$ (cf. Kellermann and Pauliny--Toth 1969).

The stationary spectrum of the PIC is obtained by solving the coupled
transport equations of secondary pairs and photons in the
high energy limit ($\nu_{\rm PIC}\gg \nu_{\rm c}$
where $\nu_{\rm c}\approx 10^{14-16}$~Hz refers to the characteristic
synchrotron frequency
of electrons at the maximum energy corresponding to equal
acceleration and cooling time scales) as
shown in MKB.  In this limit the showers develop fully linear
in the soft photon soup of the accelerated electrons,
thereby redistributing the power injected at highest
energies rather smoothly between hard X--rays and
gamma rays up to 1--10~TeV.
The shape of the spectrum results from merging of individual cascade
generations
tending to yield the
spectral index $\alpha_\gamma=1$ (equal power per decade of energy).
The upper turnover is determined by
$E_\gamma^*$ from Eq.(4) and the break towards the
X--ray spectrum with $\alpha_X=0.5-0.7$  is determined
by the characteristic synchrotron frequency of the pairs produced
by gamma rays at $E_\gamma^*$, {\it i.e.}
$E_\gamma^{**}\simeq 10^{-12}B_\perp D_{\rm j}^2l^{-2}\thinspace
{\rm MeV}$.  So $E_\gamma^{**}\simeq {\rm MeV}$ for typical
values $B_\perp=1$~G,
$D_{\rm j}=10$
and $l=10^{-5}$.

Since $a_{\rm s}<1$
for a quasi--stationary synchrotron source and since the virtual photon
scattering in the rest frame of the radiating electron is
still nonrelativistic, the emission
is optically thin (nonrelativistic) synchrotron radiation even at these
high energies.

With given values of $k_{\rm e}^{\rm BK}$,
$\eta$, $\Delta$, $\gamma_{\rm j}\approx
\Phi^{-1}$, $\theta$
and considering a source at the luminosity distance
 $D_{\rm l9}$  in Gpc
with kinetic jet power $L_{44}$ once can calculate the
observed power
$[\nu S_\nu(\nu >\nu_{\rm b}]_{12}$ in units
of $10^{12}$~Hz~Jy
$$[\nu S_\nu]_{12}=0.03k_{\rm e}\Delta^{-1}
\left(1+{2\over 3}k_{\rm e}\Lambda_{\rm e}\right)^{-1}
\gamma_{\rm j}^{-1}D_{\rm j}^{3}L_{44}D_{\rm l9}^{-2}\eqno(23)$$
While the total proton induced luminosity has the particularly
simple form Eq.(16), the cascade spectrum depends on the
radiation compactness (shifting $E_\gamma^*$)
and the magnetic field (shifting the characteristic synchrotron
frequency), which
can be calculated for a given source at $r_{\rm b}$
from
$$\eqalign{&B_{\rm b}=B_1r_{\rm b,ob}^{-1}\sin\theta\cr & =
k_{\rm e}^{-{1\over 6}}
\Delta^{7\over 12}
\left(1+{2\over 3}k_{\rm e}\Lambda_{\rm e}\right)^{7\over 12}
\gamma_{\rm j}^{13\over 6}\beta_{\rm j}^{19\over 12}
D_{\rm j}^{1\over 6}
(\sin\theta)^{{7\over 6}}
\Phi_{\rm ob} L_{\rm 44}^{-{7\over 12}}\cr}
\eqno(24)$$
and  Eqs.(18) and (23).

A striking trend follows: $l\propto
k_{\rm e}^{11\over 12}\propto (1+\eta)^{-{11\over 12}}$,
while $B_{\rm b}\propto k_{\rm e}^{
-{1\over 6}}\propto (1+\eta)^{1\over 6}$.
But from Eq.(23) one can see that
with an increasing proton/electron
ratio $\eta$ the target luminosity ($10^{9-16}$~Hz) decreases,
while -- because of Eq.(16) -- the proton induced luminosity
and the magnetic field remain roughly
constant.  Thus
the most luminous
gamma ray emitters (relative to their radio to optical
luminosity) can
emit the highest gamma ray energies in their rest frame (Fig.1).

For very low values of the
compactness $l\ll 10^{-6}$,
which should not occur in variable sources but in extended hot spots,
the spectrum changes radically, since
the individual cascade generations show up producing a bumpy
spectrum (see MKB).  For values $l\gg 10^{-3}$ the X--ray spectral
index approaches $\alpha_X=1$ and the reprocessing becomes
partially nonlinear.  Such values of $l$ occur when $r_{\rm b}$
is very small, {\it i.e.} for $r_{\rm b}\ll 1$~pc.  It must be
emphasized that thermal reprocessing (Zdziarski et al. 1990)
is not considered here because of the low thermal
gas densities in the jet (cf. Section 2.1.1).
However, for a compactness value
reaching the saturation value
the pairs produced by the cascade increase the thermal
particle density considerably (leading to a strong
annhihilation line). Then the approach of this work
is not valid anymore.

There is some heating connected to the
relativistic particles.  However, particles
in the fully ionized jet plasma rather couple to collective degrees
of freedom instead of suffering single particle Coulomb interaction.
In fact, if the latter process would yield more heat than relativistic
particle power, there would not be any acceleration.

It is important to note, however, that while electrons still
cool when they are at great distances to the acceleration regions
generating extended emission patterns like radio lobes,
protons would not behave in such a way.  Their cooling relies on
the presence of compact target photon fields.

Fig.~2 shows the variation of $\theta$ with
$\eta$ held fixed.
  It is obvious that in this case the PIC
spectrum does not change much its shape,
since the compactness is almost constant.
However, there is a positive correlation between
flux and maximum gamma ray energy (due to Doppler--boosting).
The primary electron induced spectrum, the soft spectrum,
 is interesting in that the break frequency decreases with
decreasing flux.  Thus observations showing
such a correlation would indicate  differential
Doppler--boosting.

\titleb{Application}
There are no existing data sets covering the enormous
bandwidth of the proton blazar spectrum
of more than 18 orders of magnitude
simultaneously. This makes any interpretatory attempt
of data from different epochs difficult, since
variability of the blazar emission is present on
times scales from days (due to shocks propagating
through inhomogenities) to years (due to adiabatic
expansion).  However, to demonstrate the power of
the model Figs.3 and 4 show
synthetic
spectra approximating observations of
the BL~Lac object Mkn421 and the OVV quasar 3C279
grouping existing non-simultaneous data into high--state
and low--state spectra.

The synthetic spectra do not account
for interstellar absorption both
in the host galaxy and in the Milky Way.  Moreover, the possibly
important absorption of gamma rays above 0.1~TeV for sources
at redshifts $z\ge 0.1$ on the intergalactic infrared photon
background field (Stecker et al. 1992) is not included.
This effect would qualify blazars with redshift $z<0.1$ as the
most promising TeV gamma ray candidates, {\it e.g.}
Mkn421 and similar objects like Mkn501,
3C371, Mkn180, IZw186,
4C04.77, PKS0548--322 and PKS0521--365.

The general shape of the PIC spectrum makes it clear that the GRO/EGRET
instrument is particularly well--suited to detect
proton blazars.
The gamma ray spectral index is $\alpha_\gamma\approx 1$
with a tendency for slight hardening and
the hard X-ray
spectral indices are close
to $\langle\alpha_X\rangle\approx 0.5-0.7$  consistent with the
findings by Worrall and Wilkes (1990).
The steepening towards $\alpha_\gamma\approx 1$ occurs in
the MeV range, {\it i.e.} in the GRO/COMPTEL range (cf. Section 2.2).

\tabcap{1}{Physical parameters inferred for the blazars
3C279 and Mkn421 ($H_o=75$~km s$^{-1}$Mpc$^{-1}$ and
$q_o=0.5$).  Intrinsic parameters as in Fig.1.}
\noindent
\def\quad{\hskip 1pt\relax}
{\offinterlineskip\tabskip=0pt
\hfill\vbox{\halign{\strut
\vrule#&
\hfil\quad\rm#\quad\hfil&
\vrule#&
\hfil\quad\rm#\quad\hfil&
\vrule#&
\hfil\quad\rm#\quad\hfil&
\vrule#&
\hfil\quad\rm#\quad\hfil&
\vrule#
\cr
\noalign{\bf \hrule}
&       &&    \bf 3C279 -- High && \bf --  Low&& \bf Mkn421   &\cr
\noalign{\bf \hrule}
& $z$ && $0.538$ && $0.538$  && $0.0308$ &\cr
\noalign{\bf \hrule}
& $\gamma_{\rm j}$  &&$20$ &&$20$&&$10$ &\cr
\noalign{\bf \hrule}
& $\theta$ \rm [deg] &&$3$ &&$5.5$&&$3$ &\cr
\noalign{\bf \hrule}
& $\eta$  &&$10$ &&$10$&&$1$ &\cr
\noalign{\bf \hrule}
& $k_{\rm e}$  &&$0.05$ &&$0.05$&&$1$ &\cr
\noalign{\bf \hrule}
& $L_{44}$  &&$3\times 10^3$ &&$3\times 10^3$&&$10^{-1}$ &\cr
\noalign{\bf \hrule}
& $\nu_{\rm c}$ [\rm Hz] &&$10^{14}$ &&$10^{14}$&&$10^{16}$ &\cr
\noalign{\bf \hrule}
\noalign{\bf \hrule}
& $\nu_{\rm b}$ [\rm Hz] &&$2\times 10^{11}$ &&$7.9\times 10^{10}$
&&$1.6\times 10^{13}$ &\cr
\noalign{\bf \hrule}
& $D_{\rm j}$  && $19$ && $8.5$ && $15.7$ &\cr
\noalign{\bf \hrule}
& $l$   && $2.4\times 10^{-5}$ && $2.3\times 10^{-5}$&&
$6.5\times 10^{-4}$ &\cr
\noalign{\bf \hrule}
& $B_{\rm b}$\rm [G]  &&$ 1.4 $ && $1.0$ &&90.0 &\cr
\noalign{\bf \hrule}
& $a_{\rm s}$  &&$ 4.5\times 10^{-3} $ && $4.5\times 10^{-3}$
&&$5\times 10^{-1}$ &\cr
\noalign{\bf \hrule}
& $r_{\rm b,ob}$\rm [pc]  && $5.3\times 10^{-2}$ &&  $1.7\times 10^{-1} $
&&$ 6.7\times 10^{-6}$ &\cr
\noalign{\bf \hrule}
& $\gamma_{\rm p,b}$ && $2\times 10^{11}$ && $2\times 10^{11}$ &&
$5\times 10^9$&\cr
\noalign{\bf \hrule}
& $E_\gamma^*$\rm  [TeV] &&$ 26.0 $ && $13.4$ && $0.8$ &\cr
\noalign{\bf \hrule}
& $\Phi_{\rm ob}$ \rm [deg] &&$ 38.3 $ && $20.1$ && $38.3$ &\cr
\noalign{\bf \hrule}
\noalign{\hrule}}}\hfill
}
\vskip.3cm

\titlea{Contribution of proton blazars to the
diffuse gamma ray and neutrino background}

\titleb{Diffuse gamma rays}
The acceleration of protons should be a generic property of all
radio jets.  However, as shown above, only the twofold condition
of very high proton energies and a (comoving frame)
radiation compactness greater
than $\approx 10^{-6}$ leads to significant gamma ray fluxes in the
extragalactic window $E_\gamma< 100$~TeV.  It is assumed here
that the conditions are fulfilled at least during blazar episodes
in compact radio jets.  This assumption is confirmed by the GRO
detections.  However, further gamma ray measurements must clarify the
situation.

Urry et al. (1991) give a present day space density of
$$N_{\rm BL}\approx 4\times 10^{-8}\thinspace
{\rm Mpc}^{-3}\eqno(25a)$$
for BL~Lacs at the maximum of the distribution of $N_{\rm BL}L_{\rm r}^2$
at the $2.7$~GHz luminosity $L_{\rm r}=2.7\times 10^{41}$~erg/s
and for flat--spectrum radio quasars correspondingly
$$N_{\rm FSRQ}\approx 3\times 10^{-10}\thinspace
{\rm Mpc}^{-3}\eqno(25b)$$
at $L_{\rm r}=9\times 10^{42}$~erg/s.  With an average
radio/X--ray spectral index $\langle\alpha_{\rm rx}\rangle=0.82$ one
obtains the corresponding X--ray luminosity
$L_{\rm X}\simeq 30L_{\rm r}$.  The spectral luminosity of
gamma rays from proton blazars is now estimated assuming $\eta=1$
and $\alpha_\gamma=1$ (cf. Figs.2,4) yielding
$L_E\approx L_{\rm X}E^{-1}\approx 30L_{\rm r}E^{-1}$.

\begfig6cm
\figure{3}{The OVV quasar 3C279 interpreted as a proton blazar.
Data are from the plots in Makino et al. (1989) and Hartman et al. (1992).
The synthetic spectrum is produced by shock accelerated electrons
below $10^{15}$~Hz and
analogously
by accelerated protons above this frequency.  The physical conditions
leading to the type of spectra shown are listed in Table 1.
Note that gamma rays above 0.1~TeV are likely to be absorbed
by intergalactic infrared photons.}
\endfig

\begfig6cm
\figure{4}{The Bl~Lac object Mkn421.
Data are taken from the plots in George et al.
(1988),
Michelson et al. (1992) and Punch et al. (1992), physical
parameters are listed in Table 1.  The high--state X--ray emission
could still be proton initiated emission, if thermal reprocessing is assumed,
{\it e.g.} caused by a massive screen crossing the compact jet.}
\endfig

Relevant for the diffuse flux is
the sum of the contributions of both populations BL~Lac and FSRQ,
each presumably containing a proton blazar.  This yields together
with a power--law evolutionary scheme
$$\eqalign{&\langle N_{\rm pb}L_E\rangle=30\left(N_{\rm BL}L_{\rm r}+
N_{\rm FSRQ}L_{\rm r}\right)E^{-1}
(1+z)^\beta\cr &\approx
4\times 10^{35}{\rm \thinspace erg
\thinspace s^{-1} \thinspace Mpc^{-3}}E^{-1}(1+z)^\beta\cr}
\eqno(26)$$
The  gamma ray background intensity of the proton blazars
for a Friedmann-universe with
$q_o=1/2$ and $H_o=75$~km/s /Mpc
is then given by
$$I\left(E\right)=
{3c\over 8\pi H_o}\int_0^{z_{\rm max}}
N_{\rm pb}L_{E(1+z)}(1+z)^{\beta-5/2}dz
\eqno(27)$$
where $z$ denotes redshift.
The result is only weakly dependent on
$\beta$, {\it viz.} for $\beta=3$ and $z_{\rm max}=1.5$
the intensity has the value
$$I(E)\simeq  2\times 10^{-8}
 E_{\rm GeV}^{-1}
\thinspace ({\rm cm^2
\thinspace s  \thinspace ster})^{-1}
\eqno(28)$$
This compares to the extrapolated intensity of gamma rays
from the Galactic disk
$$I_{\rm G}(E)\simeq 10^{-5}E_{\rm GeV}^{-1.7}\thinspace
{\rm (cm^2\thinspace s\thinspace ster)^{-1}}\eqno(29)$$
The two spectra cross each other at $E_\gamma\approx 7$~TeV
where the intensity is $I_{7\rm TeV}\simeq 3\times 10^{-12}
{\rm (cm^2\thinspace s\thinspace ster)^{-1}}$.
The value is well below
the quiet uncertain upper limits which can
be inferred from measurements reported in Ressell
and Turner (1990)
$$I_{\rm 20TeV}\le 10^{-11}
\left({\rm  cm^2
\thinspace s \thinspace ster}\right)^{-1}\eqno(30)$$
The intergalactic absorption predicted by
Stecker et al. (1992) reduces the flux of extragalactic
gamma rays above 0.1~TeV to the fluxes of
the individual point sources with redshifts $z<0.1$.

\titleb{Diffuse neutrinos}
Eq.(27) can be used to estimate the flux of neutrinos escaping from
the proton blazar
(from charged pion decay; kaons and neutrons can be neglected).
The neutrino spectrum is very hard $dN/dE\propto E^{-1}$,
so that its (comoving frame) luminosity  peaks at
$1/20$ of the proton maximum energy
from Eq.(15).
This is, of course, in marked contrast to the electromagnetic
spectrum where
the effect of cascading washes out the memory
of the injection energy.
The physical reason for the injection at highest energy
is that in an inverse power--law target field the highest energy protons
find the most target photons (Mannheim and Biermann 1989).

Therefore the observed spectral luminosity not too far from the maximum energy
is constant
$$L_{E_\nu}=L_XE_{\nu,\rm max}^{-1}\Lambda_\gamma\eqno(31)$$
where $\Lambda_\gamma=\ln E_\gamma^*/E_{\gamma}^{**}\approx 5$.
Normalization guarantees
$$L_{\nu}=\int L_{E_\nu}dE_\nu=L_X\Lambda_\gamma
\approx L_{\rm PIC}\eqno(32)$$
(the gamma ray spectral luminosity is lower by the factor
$\Lambda_\gamma^{-1}$ because of the cascade smearing photons over
the energy band from $E_\gamma^*$ to $E_\gamma^{**}$).
The  omnidirectional flux is then given by
$$F_\nu \approx
10^{-15}\thinspace{\rm (cm^2\thinspace
s})^{-1}  \ \
{\rm for}\ \
E_\nu <
{D_{\rm j} E_{\rm p,max}\over (1+z_{\rm max}) 20}\approx
10^{9}\thinspace {\rm GeV}\eqno(33)$$
($\gamma_{\rm p,b}\approx 10^{10}$, $D_{\rm j}\approx \langle \gamma_{\rm
j}\rangle\approx
7$ from Urry et al. 1991).
Eq.(33) compares to the flux of atmospheric neutrinos
$$F_\nu(E_\nu)\approx \left[ E_\nu\over {\rm GeV}\right]^{-2.3}
\thinspace {\rm (cm^2\thinspace s)^{-1}}\eqno(34)$$
so that the extragalactic blazar neutrino flux crosses the
atmospheric flux at an energy
$$E_\nu^\star\approx 3\ {\rm PeV}\eqno(35)$$
where the detection probability of $\bar\nu_{\rm e}$ is enhanced
due to the Glashow--Weinberg resonance at 6.4~PeV.
The possibility of detecting neutrinos
from flat--spectrum radio quasars with the Fly's Eye
experiment
is investigated in Mannheim et al. (1992).  There is a significant
chance that the  HiRes  (Cassiday et al. 1989)
might be able to detect proton blazars.
Taking into account
neutrinos from pp collisions (Section 2.1.1) the flux Eq.(33) is changed
at low energies to a $F_\nu\propto E^{-1}$ behaviour.
Since the p$\gamma$ flux is flat, the de--redshifted
energy, at which steepening due
to neutrinos from pp collisions occurs, is given by $(1+z_{\rm max})
E_{\nu\rm max}7
\left[r_{\rm S}/ r\right]$.  Assuming an average value of $r\approx
10^{3}r_{\rm S}$
(the mean value for 3C279 and Mkn421) one has steepening below
$E_\nu\approx 10^7$~GeV.  This would decrease the value of $E_\nu^*$
to $\approx 1 $~PeV.

\titlea{Leading neutrons}
Hadronic interactions exhibit a unique fingerprint:
In every second inelastic
proton-photon event the leading nucleon emerging
from the interaction fireball is a neutron.
Since neutrons are not magnetically confined anymore,
they can escape the acceleration region in the
jet.  The optical depth for neutrons transverse to
the jet
is less than unity,
so that the ultra--relativistic neutrons
can leave the source without turning back into
confined protons (provided that the distance of the
jet from the site of the thermal UV nucleus is great enough
to prevent damping by these photons).
Since the frame in which the
scattering centers isotropize the accelerated protons
rests in the bulk flow of the
jet, the emitted neutrons as seen from
an observer stationary with respect to the
host galaxy are streaming along a cone
with the blazar in its apex.
As a corollary it follows that

\noindent
{\it Baryonic blazar beams contain  energetic
photons, neutrinos and neutrons with comparable luminosities.}

It remains to be shown, wether the neutrons have observable consequences (cf.
Kirk and Mastichiadis 1989).

While it has been suggested
by Sikora et al. (1989) and
Begelman et al.
(1991)\footnote{$^{2)}$}{cf.  MacDonald
et al. (1990) for further
consequences of energetic
neutrino production.} that neutrons from
a hypothetical accretion disk in an AGN can
explain  observed gas outflows, the mechanism proposed
works best for not too energetic neutrons.  To convert
the power of the relativistic neutrons into kinetic power of
thermal plasma authors assume effective coupling of the two
media via excitation of Alfv\'en--waves by $\beta$--decay protons.

However, it is by no means clear how the microphysics shall
interplay
to lock the  particles isotropically to the plasma when their energy
is very high
({\it e.g.}, Berezinski\u i et al. 1990).
It must be remembered that the luminosity of
neutrons from the jet peaks at the highest neutron energy
($\approx 10^9$~GeV).
Without such isotropic locking there are no adiabatic losses.
Moreover, the distance neutrons travel before suffering
$\beta$-decay is given by
$$R_n=\gamma_nc\Delta\tau_n\approx
\left(\gamma_n\over 10^8\right)\thinspace
{\rm kpc}\eqno(36)$$
Proton blazars generating most neutron luminosity at
$\gamma_{\rm n}\ge 10^9$ are therefore clearly injectors
of cosmic rays, because the neutrons decay at $R_{\rm n}\ge 10$~kpc
outside the main
central galaxy.  Only for $\gamma_{\rm n}< 10^9$ neutrons
decay well within the host galaxy.   The greatest number of these neutrons
come from pp collisions in the jet (Section 2.1.1). Their luminosity is
much less than that of neutrons from p$\gamma$ collisions.  It must
also be remembered that the great luminosity of the proton blazar
appears only in the direction of the jet.  After isotropization the
true luminosity available for conversion into the kinetic power of
a wind is reduced by the factor $D_{\rm j}^{-4}\approx 10^{-4}$.

Nevertheless one can proceed and calculate the expected properties
of a neutron driven wind, assuming that
after $\beta$--decay rapid coupling to the plasma can be achieved.
The energy deposition
rate for the neutrons is then given by
$$dL_{\rm n}/dlogE_{\rm n}\simeq D_{\rm j}^{-4}L_{\rm X}\Lambda_\gamma
E_{\rm n}/
E_{\rm n,max}\eqno(37)$$
due to the hard injection $dN_{\rm n}/dE_{\rm n}\propto E_{\rm n}^{-1}$
(Mannheim and Biermann 1989).
Assuming that the neutron luminosity converts
into kinetic power at the distance of maximum energy deposition, {\it i.e.}
$$L_{\rm kin}=\dot Mc^2\beta_{\rm W}^2/2
\simeq D_{\rm j}^{-4}L_X\Lambda_\gamma \eqno(38)$$
yields together with flux conservation (constant mass flux)
$$\rho v_WA=\dot M\eqno(39)$$
where $\rho$ denotes mass density and $A$ the wind's cross
sectional area,
the wind speed
$$\eqalign{&\beta_{\rm W}\simeq  10^{-3}
\left({D_{\rm j}^{-4}L_{\rm X}\Lambda_\gamma\over 5\times 10^{39}\thinspace
{\rm erg/s}}\right)^{1/3}\left({R_{\rm n,max}\over
10\thinspace{\rm kpc}}\right)^{-2/3}
\cr &\times \left(\tan\left[\Theta\right]\over\tan\left[ 10^{\rm o}
\right]\right)^{-2/3}
\left({n_{\rm th}\over 10^{-3}\thinspace{\rm
cm^{-3}}}\right)^{-1/3}\cr}\eqno(40)$$
and mass flux
$$\eqalign{&\dot M=0.2M_{\odot}/{\rm yr}
\left({D_{\rm j}^{-4}L_{\rm X}\Lambda_\gamma
\over 5\times 10^{39}\thinspace{\rm erg/s}}\right)^{1/3}
\left({R_{\rm n,max}\over
10\thinspace{\rm kpc}}\right)^{4/3}
\cr &\times
\left(\tan\left[\Theta\right]\over \tan\left[10^{\rm o}\right]\right)^{2/3}
\left({n_{\rm th}\over 10^{-3}\thinspace{\rm
cm^{-3}}}\right)^{2/3}\cr}\eqno(41)$$
The opening angle of the cone is denoted as $\Theta$ and
the value $\Theta\approx 10^{\rm o}$ is taken from  Urry et al.  (1991).
A motion of the position angle of the compact jet with respect
to the large--scale jet may result in an effectively larger cone width.
Note that the right hand side of Eq. (38)
is a consequence of the cosmic ray energy deposition rate
$$\dot E_{\rm cr}(R)=dL_{\rm n}/dlogE_{\rm n}
(E_{\rm n})
|_{R=
E_{\rm n}c\Delta\tau_{\rm n}/m_{\rm n}c^2}
\eqno(42)$$
This conical wind surrounding the radio jet
in giant radio galaxies would have the remarkable property
$$L_{\rm kin}\propto L_{\rm nth}\propto
L_{\rm jet}\eqno(43)$$
where $L_{\rm nth}$ denotes the luminosity of the nonthermal blazar.
The blazar is, of course, barely observable in FR~galaxies where
it is beamed in the plane of the sky -- unless the
contiuum is reradiated by photoionized clouds crossing the
beam.  On the other
hand, when the radio galaxy appears as a blazar, the effect of the wind
would be difficult to disentangle, since the blueshifted line (w.r.t. the
host galaxy) emitting gas has many orders of magnitude less power.

Eq.(43) bears some similiarity with the
observational result (Rawlings and Saunders 1990)
$$L_{\rm NLR}\propto L_{\rm jet}\eqno(44)$$
where $L_{\rm NLR}$ denotes the narrow line luminosity.
The NLR can be viewed as photoionized clouds reradiating the
assumed blazar continuum.
However, Eq. (37) shows that because of the high neutron energies
most  power is deposited many kiloparsecs away from
the kinematical center of the galaxy, in contrast to the observed
NLR and  fast nuclear wind in some
FR galaxies, which are found at
only a few hundred parsecs.  At greater distances from the center
extended line emitting structures can still be
observed in many FR galaxies
(Tadhunter et al. 1989) and the neutron driven wind
would imply that these structures represent an outflow.

\titlea{Conclusions}
What can be learned from adding relativistic protons to the usual
relativistic electrons observed in compact radio jets?

First of all, if the protons manage to be accelerated up to the
high energies where their energy loss rate is equal to the electron
energy loss rate, they generically induce high energy emission from
X--rays to gamma rays.  The flux spectrum of the  polarized synchrotron
cascade radiation has an index $0.5-0.7$ at X--ray frequencies steepening
in the MeV range to a gamma ray flux with index
$0.8-1.0$ up to the TeV range.
The electromagnetic synchrotron showers in the jet
are unsaturated (only four generations of photons)
and yield neither a significant annihilation line
nor much thermal energy.

At present it is unknown, wether
there is also a generic proton/electron ratio $\eta$ and hence
a generic ratio $L(>X)/L(<UV)$.   The EGRET detections
of flat--spectrum sources seem to indicate that either there is
significant scatter in the distribution of $\eta$ or that the
proton acceleration (taking longer than the electron acceleration)
is sometimes interrupted before reaching the maximum proton energies.
Proton enrichment $\eta>1$ increases the ratio of gamma rays to
radio photons, so that it is clear, why EGRET found radio sources
below the $1$~Jy level with stronger gamma ray emission than
sources with radio fluxes above $1$~Jy.

Secondly, with protons brought into the game, the flat--spectrum
radio sources generate a diffuse background of gamma rays and
neutrinos both of which are detectable with planned experiments
above 7~TeV and 1~PeV, respectively.

Finally, the presence of protons leads to neutron production which
has remarkable astrophysical consequences.  At ultra--high energies
the neutrons escape from the host galaxy without adiabatic
losses injecting cosmic ray
protons.  At lower energies they can accelerate gas surrounding
the radio jet within the host galaxy (and its halo).
\ack{I acknowledge support by DARA grant  FKZ 50 OR 9202 and
many helpful discussions with  P.L.~Biermann. A.~Witzel
and R.~Wegner from the VLBI--group in Bonn have contributed with
their experience and expertise
about the fascinating realm of superluminal radio sources.
I also acknowledge discussions with R.~Schlickeiser
about the issue of leptonic vs. baryonic cosmic rays and
helpful comments on the manuscript by A.~Zdziarski.
}
\begref




\ref
Begelman, M.C., de~Kool, M., Sikora, M.,  1991, {\it Astrophys. J.},
{\bf 382}, 416

\ref
Bell, A.R., 1978, {\it Monthly Notices Roy. Astron. Soc.} {\bf 182}, 443

\ref
Berezinski\u i, V.S., Bulanov, S.V., Dogiel, V.A., Ginzburg, V.L. (ed.),
Ptuskin, V.S., 1990, {\it Astrophysics of Cosmic Rays}, North Holland,
Amsterdam

\ref
Biermann, P.L., Strittmatter, P.A., 1987, {\it Astrophys. J., }{\bf 322},
643

\ref
Biermann, P.L.,  1991, in: {\it   Frontiers in Astrophysics, } eds. R.
Silberberg, G. Fazio, M. Rees, Cambridge


\ref
Blandford, R.D., K\"onigl, A., 1979, {\it Astrophys. J.}, {\bf 232}, 34




\ref
Bregman, J.N., 1990, {\it Astron. Astrophys. Rev.,} {\bf 2}, 125


\ref
Burns, M.L., Lovelace, R.V.E., 1982, {\it Astrophys. J.}, {\bf 262}, 87

\ref
Cassiday, G.L., et al., 1989, in: {\it Proc. of the Workshop on
Particle Astrophysics:  Forefront Experimental Issues, held
at the University of California, Berkley, Dec 8-10}, ed. E.B. Norman,
World Scientific, Singapore, p. 259

\ref
Ellison, D.C., Jones, F.C., Reynolds, S.P., 1990, {\it Astrophys. J.},
{\bf 360}, 702

\ref
Fanaroff, B.L., Riley, J.M.,  1984, {\it Monthly Notices Roy. Astron. Soc.}
{\bf 167}, 31P


\ref
George, I.M., Warwick, R.S., Bromage, G.E., 1988, {\it Monthly
Notices Roy. Astron. Soc.} {\bf 232}, 793



\ref
Giovanoni, P.M., Kazanas, D., 1990, {\it Nature}, {\bf 345}, 319

\ref
Halzen, F., Zas, E., 1992,
{\it University of Wisconsin -- Madison},  Preprint MAD/PH/695

\ref
Hartman, R.C., et al., 1992, {\it Astrophys. J.}, {\bf 385}, L1





\ref
Kellermann, K.I., Pauliny-Toth, I.K.,  1969, {\it  Asrophys. J., }{\bf 155},
L71



\ref
Kirk, J.G., Mastichiadis, A., 1989, {\it Astron. Astrophys.}, {\bf 211},
75

\ref
K\"onigl, A., 1981, {\it Astrophys. J.}, {\bf 243}, 700




\ref
MacDonald, J., Stanev, T., Biermann, P.L., 1991, {\it Astrophys. J.},
{\bf 378}, 30

\ref
Makino et al.,  1989, {\it Astrophys. J.}, {\bf 347}, L9

\ref
Mannheim, K., Biermann, P.L., 1989, {\it Astron. Astrophys., }{\bf 221}, 211

\ref
Mannheim, K., Kr\"ulls, W.M., Biermann, P.L., 1991,
{\it Astron. Astrophys., }{\bf 251}, 723

\ref
Mannheim, K., Biermann, P.L.,  1992, {\it Astron. Astrophys.}, {\bf 253}, L21

\ref
Mannheim, K., Stanev, T., Biermann, P.L., 1992, {\it Astron. Astrophys.},
{\bf 260}, L1



\ref
Maraschi, L., Maccacaro, T. Ulrich, M.-H. (eds.), 1989,
{\it Proc. of the Workshop on BL~Lac Objects}, Lecture
Notes in Physics, Vol. 334,
Springer Verlag, Berlin, Heidelberg, New York

\ref
Michelson et al., 1992,  {\it IAU Circ.} No. 5470


\ref
Protheroe, R.J., Szabo, A.P., 1992, {\it submitted to Nature}

\ref
Punch, C.W. et al., 1992, {\it Nature}, {\bf 358}, 477

\ref
Qian, S.J., Quirrenbach, A., Witzel, A., Krichbaum, T.P., Hummel, C.A.,
Zensus, J.A., 1991, {\it Astron. Astrophys.}, {\bf 241}, 15

\ref
Quirrenbach, A., Witzel, A., Wagner, S., Sanchez--Pons, F., Krichbaum, T.P.,
Wegner, R., Anton, K., Erlens, U., H\"ahnelt, M., Zensus, J.A., Johnston,
K.J., 1991, {\it Astrophys. J.}, {\bf 372}, L71

\ref
Rachen, J.P., Biermann, P.L., 1992, {\it to be submitted to Astron.
Astrophys.}

\ref
Rawlings, S., Saunders, R., 1991, {\it Nature}, {\bf 349}, 138

\ref
Ressell, M.T., Turner, M.S.,  1990, {\it Comments Astrophys.}, Vol. 14, No. 6,
323





\ref
Sikora, M., Kirk, J.G., Begelman, M.C., Schneider, P., 1987, {\it
Astrophys. J., }{\bf 320}, L81


\ref
Sikora, M., Begelman, M.C., Rudak, B.,  1989, {\it Astrophys. J.}, {\bf 341},
L33

\ref
Stecker, F.W., De Jager, O.C., Salamon, M.H., 1992, {\it Astrophys. J.},
{\bf 390}, L49



\ref Stenger, V.J. et al. (eds.), 1992,
{\it High
Energy Neutrino Astrophysics}, Proc. of the Workshop held at the
Univ. of Hawaii
at Manoa, Honolulu, Hawaii 23--26 March 1992,
World Scientific, Singapore, in press

\ref
Svensson, R., 1987, {\it Monthly Notices Roy. Astron. Soc., }{\bf 227}, 403


\ref
Tadhunter, C.N., Fosbury, R.A.E., di Serego Aligheri, S., 1989
in: {\it Proc. of the Workshop on BL~Lac Objects}, Lecture
Notes in Physics, Vol. 334, eds. L. Maraschi, T. Maccacaro and
M.-H. Ulrich, Springer Verlag, Berlin, Heidelberg, New York, p. 79




\ref
Urry, C.M., Shafer, 1984, {\it Astrophys. J.}, {\bf 280}, 569

\ref
Urry, C.M., Padovani, P., Stickel, M., 1991, {\it Astrophys. J.},
{\bf 382}, 501





\ref
Witzel, A., 1990, in: {\it Parsec--Scale Radio Jets}, eds. J.A. Zensus,
T.J. Pearson, Cambridge University Press, p. 206

\ref
Worrall, D.M., Wilkes, B.J., 1990, {\it Astrophys. J.}, {\bf 360}, 396


\ref
Zdziarski, A.A., Ghisellini, G., George, I.M., Svensson, R.,
Fabian, A.C., Done, C., 1990, {\it Astrophys. J.}, {\bf 363}, L1

\ref
Zensus, J.A., Pearson, T.J. (eds.),  1987, {\it Proc. of a Workshop on
Superluminal Radio Sources, held at Big Bear Solar Observatory,
California, Oct 28-30, 1986}, Cambridge University
Press
\endref
\end